\documentstyle[11pt,aaspp4,flushrt]{article}

\begin{document}

\title{Polarization change due to fast winds from accretion disks}

\author{Andrei M. Beloborodov\altaffilmark{1,}\altaffilmark{2,}} 
\altaffiltext{1}{Stockholm Observatory, S-133 36, Saltsj{\"o}baden, Sweden} 
\altaffiltext{2}{Astro-Space Center of Lebedev Physical Institute, 
Profsoyuznaya 84/32, Moscow 117810, Russia}

\begin{abstract}

A fraction of the radiation produced by an accretion disk 
may be Thomson scattered by a wind
flowing away from the disk. Employing a simple plane-parallel model of 
the wind, we calculate the polarization of the scattered radiation and 
find that its sign depends on the wind velocity, $\beta=v/c$. In the case,
$0.12 < \beta < 0.78$, the polarization 
is {\it parallel} to the disk normal, i.e., it is orthogonal to the standard 
Chandrasekhar's polarization expected from accretion disks.
The velocity of an $e^\pm$ wind is likely to saturate near the equilibrium
value $\beta_*\sim 0.5$ for which the accelerating radiation pressure is 
balanced by the Compton drag. Then the change of polarization by the wind
is most pronounced.
This may help to reconcile the standard accretion disk model with the 
optical polarimetric observations of non-blazar AGNs.

\end{abstract}


\section{Introduction}

The standard model of a black hole accretion disk predicts the polarization 
of the emerging radiation. The radiation gets polarized  
due to electron scattering that strongly dominates the 
absorption opacity in luminous accretion disks. Polarization of light 
emerging from an electron plane-parallel atmosphere was calculated by
Chandrasekhar (1960) and Sobolev (1963).  
According to their results, the polarization should be
parallel to the disk surface with a maximum of 11.7 \% when the disk 
is viewed edge-on. This simple model, however, disagrees with 
observations: the optical polarization in non-blazar AGNs is typically 
$\sim$ 1 \% or less, and it tends to be parallel to the radio jet that 
is presumably perpendicular to the disk (Stockman, Moore, \& Angel 1984;
Antonucci 1992).   
The problem of accretion-disk polarization has been discussed by a number of
authors (e.g. Gnedin \& Silant'ev 1978; Loskutov \& Sobolev 1981; 
Sunyaev \& Titarchuk 1985; Phillips \& M{\'e}sz{\'a}ros 1986; 
Coleman \& Shields 1990; Laor, Netzer, \& Piran 1990; 
Matt, Fabian, \& Ross 1993; Kartje 1995; Agol \& Blaes 1996).  
In this Letter, we suggest that the original polarization of the disk
can be changed due to Thomson scattering in a mildly relativistic wind.

Gas outflows are observed in non-blazar AGNs, and evidence for mildly 
relativistic bulk velocities has been reported (Leighly et al. 1997). 
The outflows are seen on scales far exceeding the scale of the central 
source $\sim 10^{14}$ cm, and the gas dynamics in the central region is 
poorly understood. 
It is quite probable that there is a wind from the innermost region of an 
accretion disk where the bulk of the observed radiation originates.
The wind may form due to a gas outflow from a corona of the disk. 
It may also form due to an outflow of electron-positron pairs produced by 
the gamma-rays from the disk. In the latter case, the created pairs 
cool efficiently by inverse Compton scattering down to the Compton
temperature $kT_C\sim 1-10$ keV, and the cool light $e^\pm$ plasma is 
then easily pushed away by the radiation pressure. 

To investigate the effects that a wind could introduce in the pattern of 
polarization, we consider  the simplest plane-parallel model in which 
the disk is replaced by a homogeneously emitting plane and 
the wind is assumed to be a cold vertical outflow from the plane
(by "cold" we mean that the thermal motions of particles 
in the outflow are slow compared to the bulk motion).
Thomson scattering of frequency-integrated polarized radiation 
is governed by the transfer  equations written down in $\S$ 2. 
The equations  account for the relativistic aberration of light in the wind 
rest frame, which is crucial for the polarization due to scattering. 
In $\S$ ½3 and $\S$ 4 we discuss in turn the optically thin and optically 
thick winds. The results are summarized in $\S$ 5.

\section{Basic equations}

In a plane-parallel slab, the polarized radiation is represented by the 
frequency-integrated intensities $I_{l},I_{r}$, where the index $l$ refers to
the radiation polarized in the meridional plane (defined by the normal  
and the ray), and index $r$ refers to the polarization perpendicular to this
plane. The radiation is axisymmetric with respect to the normal, with the
intensity being a function of $\mu$ -- the cosine of the angle between the 
ray and the normal. Stationary Thomson transfer in 
a cold medium moving in the vertical direction with a velocity $\beta=v/c$ 
is described by the equations,
\begin{equation}
   \mu\frac{\partial I}{\partial\tau}=(1-\beta\mu)(I-S), \qquad
   \mu\frac{\partial Q}{\partial\tau}=(1-\beta\mu)(Q-R). 
\end{equation}
Here, 
$\tau$ is the scattering optical depth that the medium would have at rest,
$I=I_r+I_l$, and $Q=I_r-I_l$.
$S$ and $R$ are source functions representing the scattered radiation.

A simple way to derive the source functions is to transform the 
radiation field to the wind rest frame where Thomson scattering is coherent.
Polarization is invariant with respect to Lorentz boosts along
the normal, and the frequency-integrated intensities $I$ and $Q$ transform as
(see, e.g., Rybicki \& Lightman 1979)
$$
  I^c(\mu_c)=D^4I(\mu), \qquad  Q^c(\mu_c)=D^4Q(\mu),
$$  
$$
  \mu_c=\frac{\mu-\beta}{1-\beta\mu},
$$
where the index $c$ stands for the comoving frame, $D=\gamma(1-\beta\mu)$
is the Doppler factor, $\gamma$ is the Lorentz factor.  
When viewed from the comoving frame, the scattered radiation is represented
by the source functions, $S^c(\mu_c)$ and $R^c(\mu_c)$, given by 
Sobolev (1963).
Transforming the scattered radiation back to the laboratory frame,
$S(\mu)=D^{-4}S^c(\mu_c)$ and $R(\mu)=D^{-4}R^c(\mu_c)$, we get 
\begin{equation}
  S(\mu)=\frac{1}{D^4}
\left[I_0^c+\frac{3}{8}(3\mu_c^2-1)(I_2^c-\frac{1}{3}I_0^c+
      Q_0^c-Q_2^c)\right],
\end{equation}
\begin{equation}
  R(\mu)=\frac{9(1-\mu_c^2)}{8 D^4}
         (I_2^c-\frac{1}{3}I_0^c+Q_0^c-Q_2^c),
\end{equation}
where $I_m^c$, $Q_m^c$ ($m=0,1,2$) are the moments of the radiation field 
in the comoving frame,
$$
   I_m^c=\frac{1}{2}\int_{-1}^1 I^c(\mu_c)\;\mu_c^m\;{\rm d}{\mu_c}, \qquad
   Q_m^c=\frac{1}{2}\int_{-1}^1 Q^c(\mu_c)\;\mu_c^m\;{\rm d}{\mu_c}.
$$
$I_m^c$ are related to the corresponding moments in the lab frame, $I_m$, by
$$
  I_0^c=\gamma^2\left[I_0-2\beta I_1+\beta^2I_2\right], \qquad
  I_2^c=\gamma^2\left[\beta^2I_0-2\beta I_1+I_2\right],
$$
$$
  I_1^c=\gamma^2\left[-\beta(I_0+I_2)+(1+\beta^2)I_1\right].
$$
$Q_m$ transform in exactly the same way. Note that $Q_0^c-Q_2^c=Q_0-Q_2$.

In a 2D approach, an initially vertical wind becomes quasi-radial 
at some height and its opacity falls off. 
We consider the simplified 1D problem in which  
the wind is replaced by a slab of vertically outflowing plasma of a given 
optical depth $\tau_0$.
At the outer boundary, $\tau=0$, we assume free escape of the radiation.
At the inner boundary, $\tau=\tau_0$, we assume the existence of a source of 
limb-darkened positively polarized radiation  
(Chandrasekhar 1960) for which $I_1\approx 0.581 I_0$, $I_2\approx 0.414 I_0$.

\section{Optically thin wind}

In the case of a small optical depth, $\tau_0$, 
the optically thin approximation holds at angles satisfying the condition,
$\tau_\mu=\tau_0(1-\beta\mu)/\mu < 1$, and equations (1) yield 
the scattered radiation emerging from the slab as 
$$
   I_{sc}\approx \tau_\mu S(\mu), \qquad
   Q_{sc}\approx \tau_\mu R(\mu).
$$
The polarization degree of the scattered radiation equals 
$p_{sc}(\mu)=Q_{sc}/I_{sc}=R/S$. From equations (2) and (3) we get 
\begin{equation}
   p_{sc}(\mu)=\frac{3(1-\mu_c^2)\eta}{(8/3)+(3\mu_c^2-1)\eta},
\end{equation}
where we have introduced the parameter
\begin{equation}
   \eta=\frac{I_2^c-(I_0^c/3)+Q_0-Q_2}{I_0^c}\approx 
        \frac{I_2^c}{I_0^c}-\frac{1}{3}.
\end{equation}
The contribution to $\eta$ from $Q_0$ and $Q_2$ is $\sim 30$ times smaller 
than that from $I_2^c-I_0^c/3$, and can be neglected. 
The parameter $-1/3<\eta<2/3$ shows to what extent the radiation field is 
stretched along the normal in the rest frame of the wind, which    
is of crucial importance for the polarization of the scattered radiation 
as $p_{sc}(\mu)$ has the same sign as $\eta$ at all $\mu$.
In Figure 1 we show the dependence of $\eta$ on the wind velocity $\beta$:
$\eta$ changes sign at $\beta_1\approx 0.12$ and $\beta_2\approx 0.78$. 
It follows that the scattered radiation remains positively 
polarized if the wind is slow, $\beta<\beta_1$. The change in sign 
of the polarization happens if the wind has a velocity in the range,
$\beta_1<\beta<\beta_2$. 

In the case the outflow is comprised of a light $e^\pm$ plasma, the time-scale
to achieve Compton equilibrium with the radiation field is small and the 
wind velocity adjusts so that the radiation pressure is balanced 
by the Compton drag (see, e.g., Phinney 1982). It means that, in the comoving 
frame, the net radiation flux vanishes, i.e., $4\pi I_1^c=0$, and the equation for 
the equilibrium velocity, $\beta_*$, reads
\begin{equation}
   \beta_*(I_0+I_2)=(1+\beta_*^2)I_1,
\end{equation}
which yields $\beta_*\approx 0.52$ for the limb-darkened radiation from the 
disk. One can check that $\eta(\beta)$ has a minimum at $\beta=\beta_*$.
It follows that a wind produces the strongest negative polarization if it 
is in Compton equilibrium with the radiation field. 
This fact is illustrated by the diagram embedded in Figure 1 that shows
the angular distribution of the disk radiation in the rest frame of the wind.
A special feature of the aberration at mildly relativistic $\beta$ is 
that the radiation field is effectively "compressed" in the vertical 
direction. It is this compression that 
results in parallel polarization of the scattered radiation, and the 
compression is naturally strongest at $\beta=\beta_*$ for which the net flux 
in the comoving frame vanishes.

Whatever velocity of the wind, $\beta$, the scattered radiation has an 
extremum of polarization at $\mu=\beta$. It corresponds to $\mu_c=0$, i.e.,
to the case the ray is perpendicular to the normal in the comoving frame,
and we denote the extremum as $p_{sc}^\perp$.
In Figure 1, $p_{sc}^\perp$ is plotted against $\beta$. 
For a mildly relativistic wind, $p_{sc}^\perp\approx\eta$,
e.g., $p_{sc}^\perp=-19$ \% for $\beta=\beta_*$. In the ultrarelativistic limit, 
the aberrated radiation gets concentrated in "head-on" direction and
$p_{sc}^\perp\rightarrow 100$ \%, as discussed by Begelman \& Sikora (1987).

An observer viewing a disk with an optically thin outflow will see 
the sum of unscattered and once scattered radiation. In Figure 2 we plot 
the observed polarization as a function of the disk inclination, $\mu$, 
for the case $\tau_0=0.1$ and $\beta=\beta_*=0.52$. The optical depth is small 
and the scattering 
in the wind weakly affects the intensity of the observed radiation. 
One can see, however, that the wind changes the pattern of polarization.
This happens because the polarization extremum of the scattered radiation 
is  boosted into angles $\mu\sim 0.5$, where the original disk polarization 
is small.
Note that the calculated slab model fails at small $\mu$ as $\tau_\mu$
increases $\propto \mu^{-1}$ to $\tau_\mu=1$ where the single scattering 
approximation is not applicable. In a real wind, the scattering region 
has typical height comparable to its radius, and a detailed 2D model is
needed to get exact pattern of polarization. To illustrate importance 
of geometry, take a cylinder of height $h$ and radius $r=\sqrt{3}h$
instead of slab. Then, at $\mu>0.5$, $\tau_\mu$ changes to 
$\tau_\mu=\sqrt{3}\tau_0(1-\beta\mu)(1-\mu^2)^{-1/2}$. Combined with
the slab source functions, $S$ and $R$, this yields a rough estimate for
the resulting polarization shown by the dotted line in Figure 2.

\section{Optically thick wind}

An optically thick $e^\pm$ outflow may emerge 
in the case a fraction of the disk luminosity is emitted above 511 keV,
as hard photons are absorbed by softer X-rays to produce $e^\pm$ pairs
(the non-linear transfer of gamma-rays above a luminous disk is discussed 
in Beloborodov 1998). Compton equilibrium is established in each layer of the 
outflow, and the bulk velocity is determined by the radiation field itself 
through the equilibrium condition (6). This means that the velocity should be 
calculated self-consistently in the transfer problem. 
Comparing the Compton equilibrium transfer to the classical transfer problem
in an electron medium at rest, we note that in addition to a constant net 
flux, $4\pi I_1(\tau)=const$, we have $I_2(\tau)=const$ (one can check
this combining the first moment of the transfer equation [1] with the 
equilibrium condition [6]).
$4\pi I_2/c$ equals the radiation pressure in the vertical direction, and 
its constancy reflects the condition that the net radiative force acting on 
$e^\pm$ vanishes in the limit $\beta\rightarrow \beta_*$. 
The equilibrium velocity varies with height, and pairs 
keep $\beta\approx\beta_*$ as $\beta_*$ is a strong attractor in phase space.

We have simulated this transfer problem numerically. 
We take as initial conditions that the disk radiation is propagating
freely in the slab, then "switch on" the scattering by the flowing pairs 
at $t=0$, and follow 
the evolution of the system until a stationary solution is established.
In the calculations, we use a grid which is homogeneous in the $\cos^{-1}\mu$ 
and $\tau$ directions. The number of grid points is $N_\tau\times 
N_\mu=200\times 300$. 
The stationary solution, $I(\tau,\mu)$ and $Q(\tau,\mu)$, that we are
looking for depends only on the parameter $\tau_0$
and does not depend on the $e^\pm$ density profile, $n(\tau)$.
Therefore, we choose the simplest profile, $n(\tau)=const$. 
The chosen time step equals $\Delta t=0.5\tau_0/n\sigma_{\rm T}cN_\tau$. 

The stationary solution for $\tau_0=3$ is shown in Figures 3 and 4. 
Figure 3 displays the wind velocity.  
In the lowest layers of the slab, the equilibrium velocity is less than 
$0.5c$ since besides the disk radiation there is 
some flux in the opposite direction due to radiation being back scattered 
by the outflow.
In higher layers, the radiation gets collimated by the scattering on the 
moving pairs, and the equilibrium velocity increases to $\sim 0.7c$. 
Figure 4 shows the radiation emerging from the slab: it is
significantly anisotropic,
and it is negatively polarized with an extremum of $p\approx -14$ \% at 
$\mu\approx 0.43$.

\section{Conclusions}

We conclude that due to Thomson scattering in a wind, 
the disk radiation originally polarized perpendicular to the normal can 
acquire parallel polarization. 
This effect arises due to relativistic aberration of light in the 
wind rest frame and takes place when the wind velocity is in the range
$0.12 < \beta < 0.78$. 
Being characteristic for mildly relativistic winds, the production of
parallel polarization 
does not occur in the previously discussed scattering by highly 
relativistic 
jets (Begelman \& Sikora 1987) which produce strongly beamed radiation
with polarization perpendicular to the jet. 

An observer viewing a disk through an optically thin wind will see
the sum of the radiation from the disk and the radiation scattered in 
the wind. If the original disk radiation 
has the perpendicular Chandrasekhar-Sobolev polarization then the scattering 
in the wind tends to diminish it. The strongest effect is produced by a wind 
with velocity  $\beta=\beta_*\sim 0.5$, where $\beta_*$ corresponds to a
Compton equilibrium of the outflow with the disk radiation field. 
Such an equilibrium is likely to be established in a wind 
comprised of an $e^\pm$ plasma.
We find in this case that a modest optical depth of the wind, $\sim 0.1$,
is enough to change the pattern of the disk polarization:
The polarization degree is reduced and there appears a range of inclination 
angles with negative (parallel) polarization. 
This change may naturally explain the optical polarization observed in 
non-blazar AGNs.

We have also found that parallel polarization is produced in an optically
thick $e^\pm$ outflow which might form above a black hole accretion disk. 
The outflow is semi-relativistic, and the
emerging radiation is markedly beamed along the disk axis. 
The bulk of the disk radiation gets scattered, its original 
polarization is lost, and the emerging radiation acquires parallel 
polarization with an extremum of about $-14$ \% at an inclination angle 
of $\sim \pi/3$. 
Note that an optically thick $e^\pm$ wind also produces an annihilation
feature above 511 keV, which has a blueshift up to $\sim 2$, albeit
smeared out by the velocity gradient in the wind.

Our general conclusion is that a wind produces parallel polarization 
when it is near Compton equilibrium with the radiation field. 
This conclusion is derived for vertical winds in slab geometry
and it remains to be investigated how it is affected by deviations from
this simple model.
In particular, a magnetic field may affect the wind direction, especially
if magnetic pressure dominates the radiation pressure, and additional 
relativistic effects are important near Kerr black holes. 
A detailed 2D model should include a radial dependence 
of the disk emission and the wind density.

\bigskip

I thank C.-I. Bj{\"o}rnsson, J. Poutanen, and R. Svensson for many 
useful remarks, and the referee for helpful comments. I acknowledge 
support from the Swedish Natural Science 
Research Council and RFFI grant 97-02-16975.

\newpage
\section*{References}
{\parindent=0pt
\noindent

Agol, E., \& Blaes, O. 1996, MNRAS, 282, 965

Antonucci, R.R.J. 1992, in Testing the AGN paradigm, ed. S. Holt,
S. Neff, C.M. Urry \\ 
\hspace*{1cm}(New York), 486

Begelman, M.C., \& Sikora, M. 1987, ApJ, 322, 650

Beloborodov, A.M. 1998, MNRAS, submitted

Chandrasekhar, S. 1960, Radiative Transfer (New York: Dover)

Coleman, H.H., \& Shields, G.A. 1990, ApJ, 363, 415

Gnedin, Y.N., \& Silant'ev, N.A. 1978, SvA, 22, 325

Kartje, J.F. 1995, ApJ, 452, 565   

Laor, A., Netzer, H., \& Piran, T. 1990, MNRAS, 242, 560

Leighly, K.M., Mushotzky, R.F., Nandra, K., \& Forster, K. 1997,
ApJ, 489, L25

Loskutov, V.M., \& Sobolev, V.V. 1981, Astrofisika, 17, 535

Matt, G., Fabian, A.C., \& Ross, R.R. 1993, MNRAS, 264, 839

Phillips, K.C., \& M{\'e}sz{\'a}ros, P. 1986, ApJ, 310, 284

Phinney, S. 1982, MNRAS, 198, 1109

Rybicki G.B., \& Lightman A.P., 1979, Radiative Processes in Astrophysics
(New York: Wiley)

Sobolev, V.V. 1963, A Treatise on Radiative Transfer (Princeton: Van Nostrand)

Stockman, H.S., Moore, R.L., \& Angel, J.R.P. 1984, ApJ, 279, 485

Sunyaev, R.A., \& Titarchuk, L.G. 1985, A\&A, 143, 374

}


\begin{figure}
\plotone{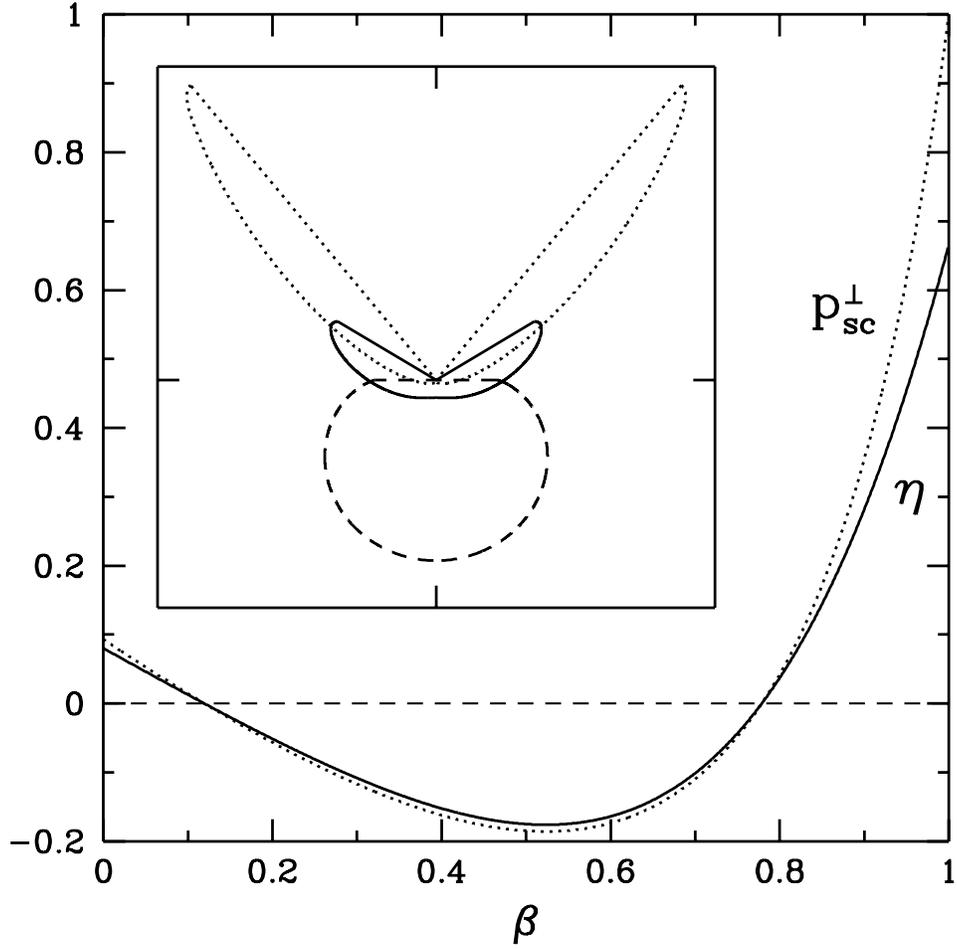}
\caption{ Dependence of parameter $\eta$ on the wind velocity, $\beta$,
(solid curve). The disk is modeled as a plane-parallel electron atmosphere 
emitting limb-darkened radiation (Chandrasekhar 1960). The dotted curve 
displays $p_{sc}^\perp$, the degree of polarization of the radiation 
scattered at right angle to the normal in the comoving frame.
The embedded diagram shows the angular distribution of the aberrated  
radiation seen by an electron in the wind, $I^c(\mu)$.
Dashed, solid, and dotted lines correspond to $\beta=0$, 
$\beta=\beta_*=0.52$, and $\beta=\beta_2=0.78$ respectively.
}
\end{figure}

\begin{figure}
\plotone{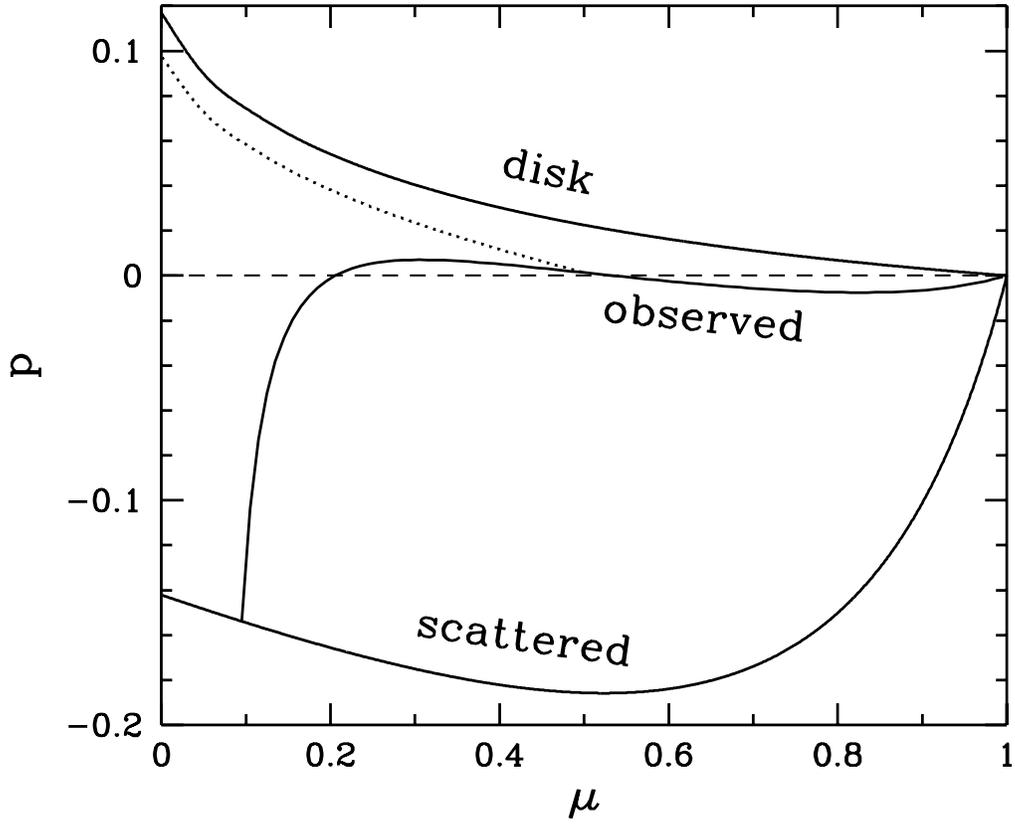}
\caption{ Angular dependence of the polarization from an 
accretion disk with a wind of optical depth $\tau_0=0.1$ and velocity
$\beta_*=0.52$. The scattered
radiation is calculated in the single scattering approximation which fails
for the slab model (solid line) when the "observed" curve approaches the 
"scattered" one at $\mu\sim 0.1$. The estimate for the cylinder model is
shown by the dotted line.
}
\end{figure}

\begin{figure}
\plotone{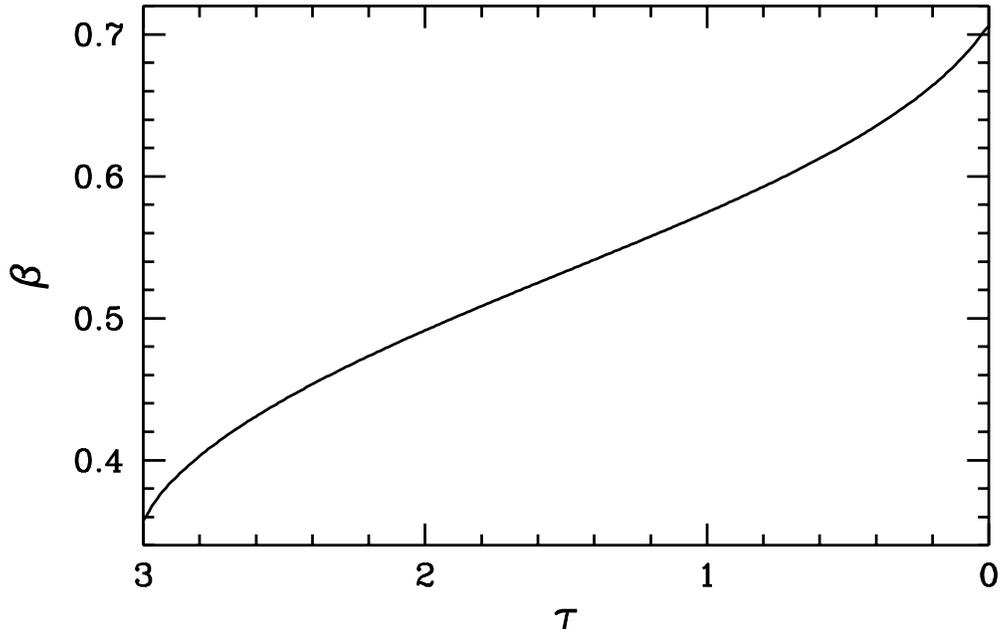}
\caption{ Velocity profile of an $e^\pm$ wind with optical depth $\tau_0=3$. 
}
\end{figure}

\begin{figure}
\plotone{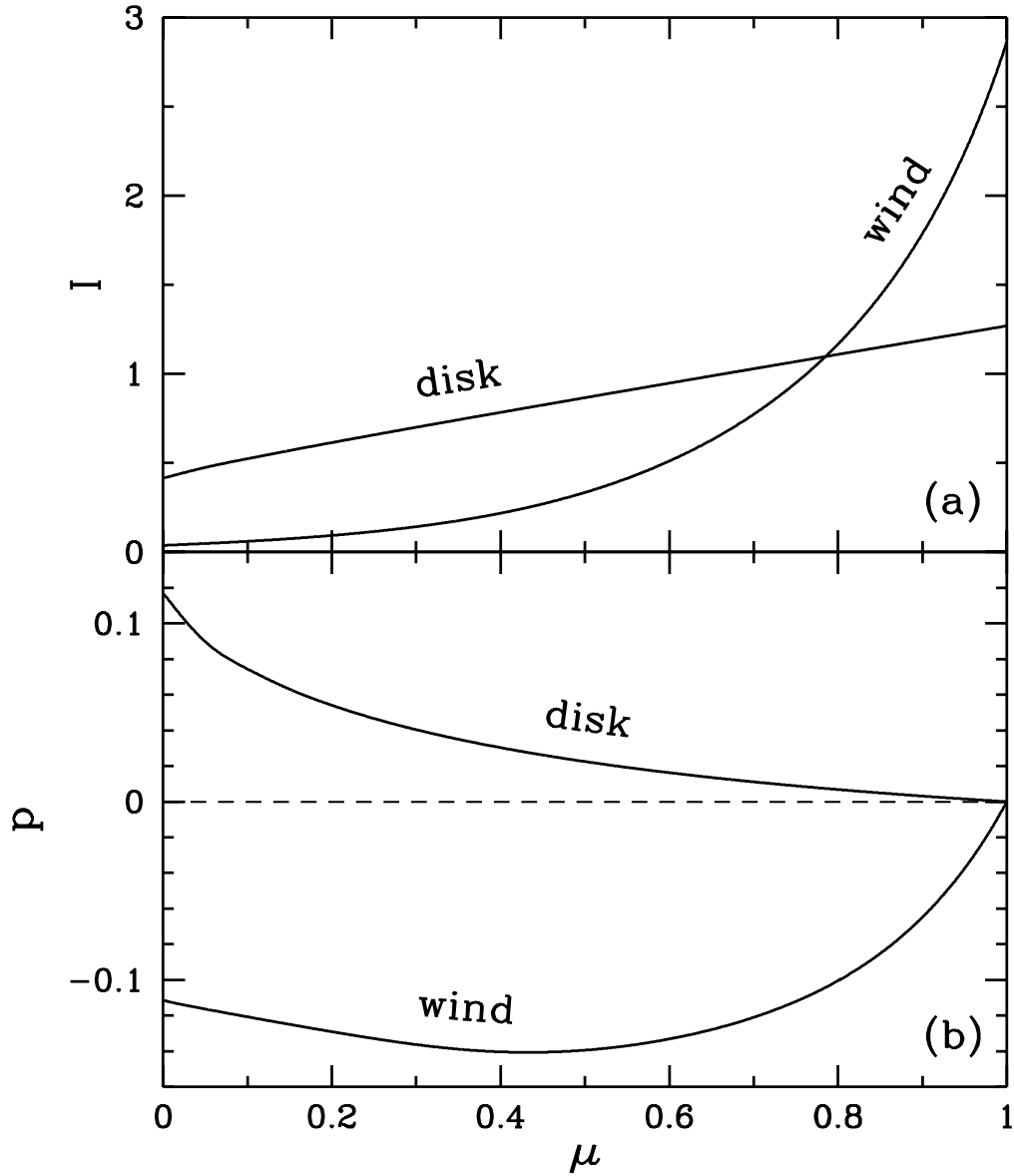}
\caption{ Radiation emerging from an optically thick $e^\pm$ wind
with $\tau_0=3$ in comparison to the Chandrasekhar's radiation of the disk:
(a) -- the angular dependence of the total intensity (normalized so that
the net flux equals $\pi$),
(b) -- the angular dependence of the degree of polarization.
}
\end{figure}

\end{document}